\documentstyle[12pt]{article}
\begin{document}
\phantom{.}
\begin{flushright}
{\bf Preprint  KOBE-FHD-96-01\\
October, 1996 }
\end{flushright}
\topmargin=0.6cm
\vsize=24.5cm
\hsize=15.5cm
\oddsidemargin=0.5cm
\vskip 2.5cm

\begin{center}

{\large\bf Anomalous Quark Chromomagnetic Moment  \\
 Induced by Instantons
\footnote{ Talk presented at the XIII International Seminar  on High
Energy Physics
 Problems,
2-7 September, Dubna,
Russia}} \\[1cm]

{\bf N.I.~Kochelev}\\

{\it Bogoliubov Laboratory of Theoretical Physics,
Joint Institute  for Nuclear Research, 141980  Dubna,
Moscow Region, Russia}
\end{center}

\begin{abstract}

 It is shown that the quark--gluon interaction induced by instantons
 leads to the quark anomalous chromomagnetic moment.
 In the instanton-liquid model for the  QCD vacuum, the chromomagnetic moment
 is $\mu_a=-\frac{\pi f}{2\alpha_s}$, where $f$ is the packing fraction
 of instantons in the vacuum.
\end{abstract}

\newpage

 A  possible  existence of the large quark anomalous
 chromomagnetic moment (QACHM) is widely under discussion in these days~
\cite{a1}--\cite{a4}.
 In~\cite{a1}, an assumption on the existence of the QACHM was made to explain
 the observed anomalies in the cross section for  reactions
with  polarized particles.
 In~\cite{a2}, the QACHM was introduced to improve the description of the
 spectroscopy of  hadrons including heavy quarks.
 It was shown in~\cite{a3} that the QACHM allows us to explain the large
 value of the  cross section for $t\bar t$ productions near threshold
 and $R_c$, $R_b$ ratio in the $Z^0$ peak.
 A very interesting explanation~\cite{a4}, based on
 the QACHM, for the excess of the
 jet cross sections at high $E_\bot$ observed by CDF~\cite{CDF} was proposed.

 It should be mentioned that, in perturbative QCD, the appearence
 of the QACHM is related with higher  order $\alpha_s$ corrections,
 and, therefore, its value needs to be small.

 In this paper, the quark anomalous chromomagnetic moment is calculated in
 nonperturbative QCD by using the liquid instanton model for
 the QCD vacuum~\cite{a5}.

 The effective Lagrangian induced by instantons has the following
 form~\cite{a6}
 \footnote{The Lagrangian (\ref{e1}) has been obtained for the case when the
 scale of the gluon-field fluctuations was significantly larger than
the size of
 the instanton, i.e., for $q<<1/\rho$, where $q$ is the gluon virtiality.}
\begin{eqnarray}
{\cal L}_{eff}&=&\int\prod_q(m_q\rho-2\pi^2\rho^3\bar q_R(1+\frac{i}{4}
\tau^aU_{aa^\prime}\bar\eta_{a^\prime\mu\nu}\sigma_{\mu\nu})q_L)
\nonumber\\
&\cdot &exp^{-\frac{2\pi^2}{g}\rho^2U_{bb^{\prime}}
\bar\eta_{b^\prime\gamma\delta}
G^b_{\gamma\delta}}
\frac{d\rho}{\rho^5}d_0(\rho)d\hat{o}
+R\longleftrightarrow L,
\label{e1}
\end{eqnarray}
 where $\rho$ is the instanton size, $\tau^a$ are the matrices of the
 $SU(2)_c$ subgroup of the $SU(3)_c$ colour group,
 $d_0(\rho)$ is the density of the instantons, $d\hat{o}$ stands
 for integration over the instanton orientation in colour space,
$\int d\hat{o}=1$,
$U$ is the orientation matrix of the instanton,
 $\bar\eta_{a\mu\nu}$ is the numerical t'Hooft symbol and
 $\sigma_{\mu\nu}=[\gamma_\mu,\gamma_\nu]/2$.

 The effective quark--gluon vertex can be obtained by expanding (\ref{e1})
 in the set of the powers of the gluon strength and by integrating over
 $d\hat{o}$\footnote{To integrate over $d\hat{o}$, one can use the
 method of~\cite{a7}.}.
 In the first order of $G_{\gamma\delta}^b$, we found out that
\begin{equation}
\Delta L_A=\sum_q\int d\rho\frac{n_{eff}(\rho)
\pi^4\rho^4}{m_q^*g}i\bar q\sigma_{\mu\nu}t^a qG_{\mu\nu}^a,
\label{e2}
\end{equation}
 where $n_{eff}(\rho)=d_0(\rho)\prod_q(m_q^*\rho)/\rho^5$ is the effective
 instanton density in the QCD vacuum,
 $m_q^*=m_q-2\pi^2\rho^2<0\mid \bar qq\mid 0>/3$ is the effective quark
 mass \cite{a5} and $t^a=\lambda^a/2$ are $SU(3)_c$ matrices.

 It should be mentioned that the Lagrangian (\ref{e2})
 was derived as a result
 of factorization of the Lagrangian (\ref{e1}).
 The factorization procedure is consisted with the contraction of some
 quark legs from the instanton vertex through the quark condensate
(see~\cite{koch} and the first reference in~\cite{a5}).

 In the instanton liqiud model for the QCD vacuum, the effective density is
\begin{equation}
n_{eff}(\rho)=n_c\delta(\rho-\rho_c),
\label{e3}
\end{equation}
 where $\rho_c$ is the average size of the instanton  and
 $n_c$ is determined by the value of the gluon condensate
 $n_c=<0\mid \alpha_sG_{\mu\nu}^a G_{\mu\nu}^a\mid 0>/16\pi$.

 Using (\ref{e3}) we obtain
\begin{equation}
\Delta L_A=
i\frac{f\pi}{2\alpha_s(\rho_c)}
\sum_q\frac{g}{2m_q^*}\bar q\sigma_{\mu\nu}
t^a qG_{\mu\nu}^a,
\label{e4}
\end{equation}
 where $f=n_c\pi^2\rho^4$ is the so-called packing fraction of instantons
 in the vacuum.

 The definition for the anomalous chromomagnetic moment is (see~\cite{a3})
\begin{equation}
\Delta L_A=-i\mu_a
\sum_q\frac{g}{2m_q}\bar q\sigma_{\mu\nu}
t^a qG_{\mu\nu}^a.
\label{e5}
\end{equation}
 Therefore, the  QACHM is given by the formula
\begin{equation}
\mu_a=-\frac{f\pi}{2\alpha_s(\rho_c)}.
\label{a6}
\end{equation}
 It should be stressed that the final result for the QACHM includes the
 value of the quark--gluon coupling constant in the {\it denominator}.
 Therefore, it is impossible to get the similar result from perturbative QCD.

 For the numerical calculation, we use the NLO approximation for the strong
 coupling constant
\begin{equation}
\alpha_s(\rho)=-\frac{2\pi}{\beta_1t}(1+\frac{2\beta_2log t}{\beta_1t}) ,
\label{e17}
\end{equation}
 where
\begin{equation}
\beta_1=-\frac{33-2N_f}{6},\ \beta_2=-\frac{153-19N_f} {12}\nonumber
\end{equation}
 and
\begin{equation}
t=log(\frac{1}{\rho^2\Lambda^2}+\delta).
\label{e18}
\end{equation}
 In Equation (\ref{e18}), the parameter $\delta\approx1/\rho_c^2\Lambda^2$
 provides a smooth interpolation of the value of  $\alpha_s(\rho)$
 from the perturbative $(\rho\rightarrow0)$ to the nonperturbative
 region \\ $(\rho\rightarrow\infty)$~\cite{a8}.

 For $N_f=3$, $\Lambda=230$MeV, $\rho_c=1.6$GeV$^{-1}$,
 by using the standard value for the gluon condensate
\begin{equation}
<0\mid \frac{\alpha_s}{\pi}G_{\mu\nu}^a G_{\mu\nu}^a\mid 0>=0.012 {\rm GeV}^4,
\nonumber
\end{equation}
 we have the estimate
\begin{equation}
\mu_a=-0.2 \ for \ \rho_c=1.6 {\rm GeV}^{-1}.
\nonumber
\end{equation}

 Unfortunately, the average size of  instantons is not well fixed
 in the model~\cite{a5}.
 For example, if one  use the same model to calculate  the masses of
 the pseudoscalar nonet  mesons based on
QCD sum rules method (see~\cite{a5}),
 the more suitable value for average size of the instantons will be
 $\rho_c=2 GeV^{-1}$.
 In this case, the value of the QACHM will be larger
\begin{equation}
\mu_a=-0.4 \ for\  \rho_c=2 GeV^{-1}.
\nonumber
\end{equation}
 In any case, the instanton contribution to the QACHM is quite large.

 From the above observation, we can say that the quark--gluon
 interaction through the QCD vacuum
 leads to the rise of the quark anomalous chromomagnetic moment.

 The applications of this result to the hadron spectroscopy and to the
 high-energy lepton--hadron and hadron--hadron scattering  are in progress.

 The author is sincerely thankful to  M.~Anselmino, A.M.~Baldin,
A.E.~Dorokhov,
 J.~Ellis, R.N.~Faustov, S.B.~Gerasimov,
E.~Leader, P.~Mulders, M.~Oka, P.~Ratcliffe,
and E.V.~Shuryak
 for useful discussions, to  T.~Morii for warm hospitality at Kobe
 University, where the part of the work was done,  and to the JSPS
 and the Heisenberg--Landau program for financial support.

\end{document}